\documentclass[11pt]{article}

\usepackage[dvipdfm,
		colorlinks=true,
		linkcolor=red,
		urlcolor=black,
		bookmarks=true,
		bookmarksnumbered=true,
		bookmarkstype=toc]{hyperref}
\usepackage[a4paper,text={6in,9in},centering]{geometry} 
\usepackage{amsmath}
\usepackage{amssymb}
\usepackage{amsbsy}
\usepackage{mathrsfs}
\usepackage{graphicx}
\usepackage{color}

\newcommand{\1}{\mbox{1}\hspace{-0.25em}\mbox{l}}

\title{%
	\vskip -3em
	\begin{flushright}{\small KOBE-TH-08-09}\end{flushright}
	\vskip 5em
	\textbf{\large Particle Propagation on a Circle with a Point Interaction}
}
\author{
Satoshi Ohya\\
\vspace{-1ex}\\
\textit{\small Graduate School of Science, Kobe University}\\[-.3em]
\textit{\small 1-1 Rokkodai, Nada, Kobe 657-8501, Japan}\\
\texttt{\small e-mail:\href{mailto:ohya@kobe-u.ac.jp}{ohya@kobe-u.ac.jp}}\\
\vspace{1ex}\\
Makoto Sakamoto\\
\vspace{-1ex}\\
\textit{\small Department of Physics, Kobe University}\\[-.3em]
\textit{\small 1-1 Rokkodai, Nada, Kobe 657-8501, Japan}\\
\texttt{\small e-mail:\href{mailto:dragon@kobe-u.ac.jp}{dragon@kobe-u.ac.jp}}
}
\date{%
\small (Dated: \today)
}
\begin{document}
\maketitle
\begin{abstract}
We study a particle propagation on a circle in the presence of a point interaction.
We show that the one-particle Feynman kernel can be written into the sum of reflected and transmitted trajectories which are weighted by the elements of the $n$-th power of the scattering matrix evaluated on a line with a point interaction.
As a by-product we find three-parameter family of trace formulae as a generalization of the Poisson summation formula.
\end{abstract}
\newpage

\section{Introduction} \label{sec:intro}
Quantum system restricted on a bounded domain has become more relevant for theoretical physics.
There, the role of boundary conditions are very important not only for the long distance (infrared) regime but also for the short distance (ultraviolet) regime.
Mathematically, the correct framework to treat the boundary conditions in quantum theory is by means of the analysis of von Neumann's self-adjoint extension of the Hamiltonian operator \cite{Reed:1975}.
Physically speaking, the variety of boundary conditions provided by the self-adjoint extension of the Hamiltonian implies that the very rich structure of point interactions available in quantum theory.

The analysis of self-adjoint extension of the Hamiltonian, as the name suggests, is essentially based on the Hamiltonian operator approach.
However, in the Feynman's path-integral approach, we do not know \textit{a priori} how to incorporate the boundary conditions into the integration measure nor into the path-integral weight.
As discussed in many textbooks (see for example \cite{Schulman:1981,Chaichian:2001}), the naive path-integral representation for a system on a bounded domain leads to a wrong boundary behavior and hence requires modification.
The most rigorous way to incorporate the boundary conditions into the Feynman kernel is to evaluate it by the operator formalism.
However, the kernel evaluated by the operator formalism becomes the summation over the energy spectrum.
In order to switch to the path-integral description, we have to perform resummation of the energy spectrum to the paths of the space.
In general, this resummation is accomplished by \textit{trace formulae}.
Trace formulae provide a direct connection between quantum energy spectrum and classical length spectrum (periodic orbits).
However, this connection is in general an asymptotic relation valid for large wave numbers, just as in the case of the Gutzwiller trace formula \cite{Gutzwiller:1971}.
There are only few cases where the trace formulae become identities.
Noteworthy among these are the Poisson summation formula and the Selberg trace formula \cite{Selberg:1956}, the former is the trace formula for the Laplace operator on flat tori and the latter on Riemannian manifolds with constant negative curvature.

Although in the operator formalism point interactions have been extensively discussed in the literature, the path-integral description of point interactions has not been fully understood yet.
In mathematically speaking, this is mainly due to the lack of trace formulae suitable for the point interactions.
In physically speaking, on the other hand, this is mainly due to the lack of our knowledge about the classical trajectories for a particle in the presence of point interactions.
The aim of this paper is to fulfill a gap of the description for boundary conditions between the operator formalism and the path-integral formalism:
we would like to propose a physically transparent prescription how to incorporate the boundary conditions obtained in the operator formalism into the path-integral description.
To illustrate our idea in a simple setting, in this paper we will consider a one-particle quantum mechanics on a circle in the presence of a single point interaction.

To begin with, let us first consider a quantum particle on a circle of circumference $L$ in the presence of a $\delta^{\prime}$-interaction described by the Hamiltonian $H = -\mathrm{d}^{2}/\mathrm{d}x^{2} + 2c\delta^{\prime}(x)$, where $c\in\mathbb{R}$ is the dimensionless coupling constant and prime ($\prime$) indicates the derivative with respect to $x$.
(Here as in the following we are using units where $\hbar = 2m = 1$.)
It is known that the $\delta^{\prime}$-interaction belongs to the so-called scale-independent subfamily of point interactions \cite{Albeverio:1998} and is verified by the boundary conditions  $\psi(L) = [(1-c)/(1+c)]\psi(0)$ and $\psi^{\prime}(L) = [(1+c)/(1-c)]\psi^{\prime}(0)$ \cite{Griffiths:1993,Kurasov:1994}, where $\psi$ is the square integrable wave function on an interval $(0,L)$.
Although the Feynman kernel of this system has been analyzed in the literature \cite{Fulop:1999,Fulop:2003}, the physical interpretation for the weight factors (see below) remains open.
We would like to first address this issue.

As ubiquitous in the scale-independent point interactions, in this $\delta^{\prime}$-interaction case the wave numbers are quantized in an integer step so that it is easy to rewrite the Feynman kernel $K(x,T; x_{0},0) = \langle x|\mathrm{e}^{-iHT}|x_{0}\rangle$ evaluated in the operator formalism into the path-integral representation with the help of the Poisson summation formula.
The resultant kernel takes the form
\begin{align}
K(x,T; x_{0},0)
&= 	\frac{1}{\sqrt{4\pi iT}}\sum_{n\in\mathbb{Z}}
	\left\{
	\cos(n\theta)\mathrm{e}^{i\frac{T}{4}\left(\frac{nL + x - x_{0}}{T}\right)^{2}}
	\mp
	\sin(n\theta)\mathrm{e}^{i\frac{T}{4}\left(\frac{(n+1)L - x - x_{0}}{T}\right)^{2}}
	\right\}, \label{eq:1-1}
\end{align}
where $0\leq\theta := \mathrm{Arccos}[(1-c^{2})/(1+c^{2})]<\pi$ and $-$ ($+$) sign for $c>0$ ($c<0$).
$\mathrm{Arccos}$ is the principal value of the inverse cosine.
Notice that the presence of a point interaction breaks the global translational invariance.
As a consequence the kernel \eqref{eq:1-1} is the sum of partial amplitudes for the translational invariant and variant classes of trajectories, which are weighted by the factors $\cos(n\theta)$ and $\sin(n\theta)$ respectively.

\begin{figure}[t]
	\begin{center}
		\begin{tabular}{cccc}
		\includegraphics{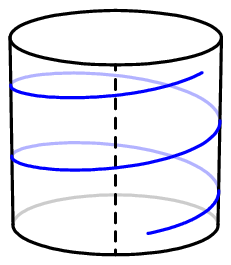} &
		\includegraphics{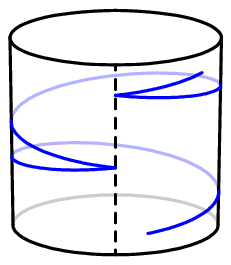} &
		\includegraphics{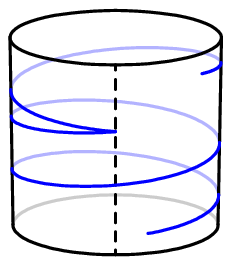} &
		\includegraphics{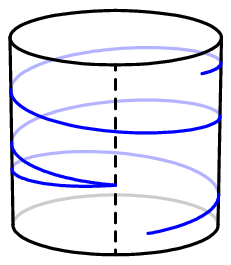} \\
		(a) $T_{+}T_{+}$ &
		(b) $R_{-}R_{+}$ &
		(c) $R_{+}T_{+}$ &
		(d) $T_{-}R_{+}$ \\
		\includegraphics{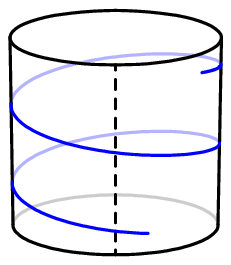} &
		\includegraphics{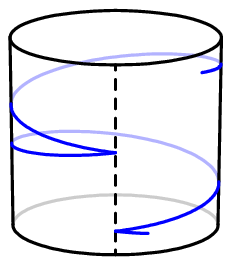} &
		\includegraphics{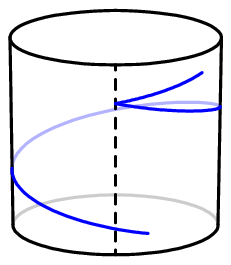} &
		\includegraphics{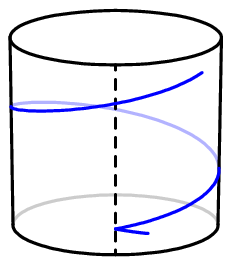} \\
		(e) $T_{-}T_{-}$ &
		(f) $R_{+}R_{-}$ &
		(g) $R_{-}T_{-}$ &
		(h) $T_{+}R_{-}$
		\end{tabular}
		\caption{Classical world lines for a particle scattered twice by the point interaction. Time is flowing along the vertical direction. The dashed line represents the world line for the point interaction. In the case of $\delta^{\prime}$-interaction with $c>0$, these $2\times2^{2}=8$ trajectories are weighted by the factors $T_{\pm}T_{\pm} = \cos^{2}\theta$, $R_{\mp}R_{\pm} = -\sin^{2}\theta$, $R_{\pm}T_{\pm} = \mp\sin\theta\cos\theta$ and $T_{\mp}R_{\pm} = \mp\cos\theta\sin\theta$.}
		\label{fig:world_lines}
	\end{center}
\end{figure}

Before going to discuss the physical meaning of the weight factors, we have to reveal the particle propagations described by \eqref{eq:1-1}.
To this end, it should first be noted that $c=0$ leads to $\theta=0$ so that Eq.\eqref{eq:1-1} becomes the well-known form of the one-particle Feynman kernel on a circle with periodic boundary conditions.
As discussed in many textbooks (see for example Ref.\cite{Schulman:1981,Chaichian:2001}), in this $c=0$ case the kernel \eqref{eq:1-1} is the sum of partial amplitudes for transitions via classical paths distinguished by the homotopy class of $S^{1}$, i.e., the winding number.
For nonzero $c$, however, the classical trajectories of a particle are not so trivial due to the presence of $\delta^{\prime}$-potential, which acts as a point scatterer.
When a particle reaches to the position of the point scatterer, there must be in general two possibilities: reflection or transmission.
Thus the paths for a particle interacting $n$-times to the point interaction must consist of $2^{n}$ distinct paths.
As an example the classical world lines for $n=2$ and $-2$ in \eqref{eq:1-1} are depicted in Figure \ref{fig:world_lines}.
As we will see below, the half of these $2^{n}$ paths belongs to the translational invariant class and the other half of them to the translational variant class.

Now it is time to discuss the physical meaning of the weight factors.
It is intuitively clear that reflected trajectory should be weighted with a reflection coefficient $R_{+}$ ($R_{-}$) for every time when a particle is reflected by the point scatterer from left to left (right to right), where $R_{+}$ ($R_{-}$) is the reflection coefficient for a particle propagating the negative half-line $\mathbb{R}_{-}$ (positive half-line $\mathbb{R}_{+}$).
Similarly, transmitted trajectory should be weighted with a transmission coefficient $T_{+}$ ($T_{-}$) for every time when a particle is transmitted by the point scatterer from left to right (right to left), where $T_{+}$ ($T_{-}$) is the transmission coefficient for a particle propagating from $\mathbb{R}_{-}$ to $\mathbb{R}_{+}$ and vice versa.
The physical meaning of the weight factors is now obvious: these must be elements of the $n$-th power of the one-particle scattering matrix $S^{(1)}$, which we would like to call the \textit{$n$-times scattering matrix} $S^{(n)}$, evaluated on a line with a point interaction at the origin (see Section \ref{sec:S-matrix}).
In the case of $\delta^{\prime}$-interaction, it is easy to compute the one-particle scattering matrix.
The result is
\begin{align}
S^{(1)}
= 	\begin{pmatrix}
	T_{+} 	& R_{-} \\
	R_{+} 	& T_{-}
	\end{pmatrix}
= 	\begin{cases}
	\begin{pmatrix}
	\cos\theta 	& \sin\theta \\
	-\sin\theta 	& \cos\theta
	\end{pmatrix}, 	& \text{for } c>0, \\
	\begin{pmatrix}
	\cos\theta 	& -\sin\theta \\
	\sin\theta 	& \cos\theta
	\end{pmatrix}, 	& \text{for } c<0,
	\end{cases} \label{eq:1-2}
\end{align}
which is just the rotational matrix.
Thus the $n$-times scattering matrix is given by just replacing the argument $\theta$ in \eqref{eq:1-2} to $n\theta$.
These matrix elements are nothing but the weight factors in \eqref{eq:1-1}.
In this sense the identity $\cos^{2}(n\theta) + \sin^{2}(n\theta) = 1$ is a consequence of the unitarity of the scattering matrix and can be viewed as the \textit{partial amplitude unitarity}.

So far we have studied only the case of $\delta^{\prime}$-interaction, it seems that the above discussion is valid for any one-particle quantum mechanics on a circle with a single point interaction.
As we will show in the rest of this paper this observation is indeed true.
Now it is time to give an explicit statement for the purpose of this paper.
The main goal of this paper is to show the following statement:
{\it
the Feynman kernel for a spinless particle moving freely on a circle of circumference $L$ with a single point interaction at the origin can be written into the following generic form:
\begin{align}
& 	K(x,T; x_{0},0)
= 	\int_{-\infty}^{\infty}\frac{\mathrm{d}p}{2\pi}
	\mathrm{e}^{iT\left[p\left(\frac{x - x_{0}}{T}\right) - p^{2}\right]} \nonumber\\
& 	+
	\sum_{n=1}^{\infty}\int_{-\infty}^{\infty}\frac{\mathrm{d}p}{2\pi}
	\left\{
	S^{(n)}_{++}(p)
	\mathrm{e}^{iT\left[p\left(\frac{nL + x - x_{0}}{T}\right) - p^{2}\right]}
	+
	S^{(n)}_{-+}(p)
	\mathrm{e}^{iT\left[p\left(\frac{(n+1)L - x - x_{0}}{T}\right) - p^{2}\right]}
	\right\} \nonumber\\
& 	+
	\sum_{n=1}^{\infty}\int_{-\infty}^{\infty}\frac{\mathrm{d}p}{2\pi}
	\left\{
	S^{(n)}_{--}(p)
	\mathrm{e}^{iT\left[-p\left(\frac{-nL + x - x_{0}}{T}\right) - p^{2}\right]}
	+
	S^{(n)}_{+-}(p)
	\mathrm{e}^{iT\left[-p\left(\frac{(-n+1)L - x - x_{0}}{T}\right) - p^{2}\right]}
	\right\} \nonumber\\
& 	+ (\text{bound state contribution}), \label{eq:1-3}
\end{align}
where $S^{(n)}_{\pm\pm}$ and $S^{(n)}_{\pm\mp}$ are the elements of $n$-times scattering matrix.}

This statement is based on the following observations:
\begin{enumerate}

\item The classical trajectories $x_{\mathrm{cl}}(t)$ for a particle propagating from $(x_{0}, 0)$ to $(x, T)$ scattered $n$-times by the point interaction are exhausted by $x_{\mathrm{cl}}(t) = x_{0} + v_{\mathrm{cl}}t$, where $v_{\mathrm{cl}} = (\pm nL + x - x_{0})/T$ for translational invariant class and $v_{\mathrm{cl}} = ((\pm n+1)L - x - x_{0})/T$ for translational variant class of classical trajectories, where `$+$' sign is for the trajectory of right-moving outgoing particle and `$-$' sign is for that of left-moving outgoing particle.

\item Any paths for a particle traveling from $(x_{0}, 0)$ to $(x, T)$ with momentum $p$ are categorized into the four cases, that is, the propagation from left to right, from left to left, from right to left, and from right to right.
The corresponding plane waves are $\mathrm{e}^{ip(nL + x - x_{0})}$, $\mathrm{e}^{ip((n+1)L - x - x_{0})}$, $\mathrm{e}^{-ip(-nL + x - x_{0})}$ and $\mathrm{e}^{-ip((-n+1)L - x - x_{0})}$, respectively.
These four classes of classical trajectories should be weighted by the factors $S^{(n)}_{++}(p)$, $S^{(n)}_{-+}(p)$, $S^{(n)}_{--}(p)$ and $S^{(n)}_{+-}(p)$, which are the elements of $n$-times scattering matrix $S^{(n)}(p)$ in the basis of right- and left-moving momentum mode (see Section \ref{sec:S-matrix}).

\item The bound state contribution, even if it exists, does not affect the scattering process on a line such that it can be added at the end of computation.
\end{enumerate}

The purpose of this paper is to show the validity of \eqref{eq:1-3} for allowed point interactions in quantum mechanics, which can be classified, as mentioned before, by means of the analysis of the self-adjoint extension of the Hamiltonian operator.
In physical language, the self-adjoint extension of the Hamiltonian is translated into the requirement for the global conservation of the probability current density $j(0) = j(L)$, where $j = -i((\psi^{\ast})^{\prime}\psi - \psi^{\ast}\psi^{\prime})$ with $\psi$ being the wave function on the Hilbert space consisting of square integrable functions on the interval $(0,L)$.
Quantum mechanical system for a free particle on a circle is known to admit a $U(2)$ family of distinct point interactions characterized by the boundary conditions \cite{Fulop:1999,Fulop:2003}
\begin{align}
 	(U - \1)
	{\vec \Psi}(0_{+})
	+ iL_{0}(U + \1)
	{\vec \Psi}^{\prime}(0_{+})
= 	\vec{0}, \label{eq:1-4}
\end{align}
where
\begin{align}
{\vec \Psi}(x)
:= 	\begin{pmatrix}
	\psi(x) \\
	\psi(L-x)
	\end{pmatrix}, \quad
{\vec \Psi}^{\prime}(x)
= 	\begin{pmatrix}
	\psi^{\prime}(x) \\
	- \psi^{\prime}(L-x)
	\end{pmatrix}, \quad
0< x<L. \label{eq:1-5}
\end{align}
$U$ is a $2\times2$ unitary matrix and $L_{0}$ is an arbitrary real constant length scale, which is just introduced to adjust the length dimension of the equation.
For the following discussions it is convenient to parameterize the matrix $U\in U(2)$ as the following spectral decomposition form:
\begin{align}
& 	U
= 	\mathrm{e}^{i\alpha_{+}}P_{+}
	+ \mathrm{e}^{i\alpha_{-}}P_{-}, \label{eq:1-6}\\
& 	P_{\pm}
= 	\frac{\1\pm\vec{e}\cdot\vec{\sigma}}{2}, \label{eq:1-7}
\end{align}
where $\vec{\sigma} = (\sigma_{1}, \sigma_{2}, \sigma_{3})$ is a vector of the Pauli matrices, $\mathrm{e}^{i\alpha_{\pm}}$ ($0\leq\alpha_{\pm}<2\pi$) are the two eigenvalues of the unitary matrix $U$ and $P_{\pm}$ is the corresponding projection operators fulfilling $P_{+} + P_{-} = \1$, $(P_{\pm})^{2} = P_{\pm}$ and $P_{\pm}P_{\mp} = 0$.
$\vec{e}=(e_{x}, e_{y}, e_{z})$ is a real unit vector satisfying the condition $e_{x}^{2} + e_{y}^{2} + e_{z}^{2} = 1$.
In this paper we derive analytical forms of the one-particle Feynman kernel with these parameters.

It is worthwhile to point out here that if we multiply the projection operators $P_{\pm}$ to \eqref{eq:1-4} on the left, the boundary conditions boil down to the following two independent equations
\begin{align}
	P_{\pm}
	\left[
	{\vec \Psi}(0_{+})
	+ L_{\pm}
	{\vec \Psi}^{\prime}(0_{+})
	\right]
&= 	\vec{0}, \label{eq:1-8}
\end{align}
where
\begin{align}
L_{\pm}
&:= 	L_{0}\cot(\alpha_{\pm}/2). \label{eq:1-9}
\end{align}
It should be noted that \eqref{eq:1-8} is not well-defined when $\alpha_{\pm} = 0$.
We will, however, use \eqref{eq:1-8} instead of \eqref{eq:1-4} as the boundary conditions by taking a careful limit for the case of $\alpha_{\pm} = 0$.

The rest of this paper is organized as follows.
In Section \ref{sec:RT} we derive the general forms of the reflection and transmission coefficients for a particle on a whole line in the presence of a point interaction at the origin.
In Section \ref{sec:S-matrix} we define the one-particle scattering matrix on $\mathbb{R}\setminus\{0\}$ and then introduce the $n$-times scattering matrix.
Section \ref{sec:spectrum} is devoted to detailed analysis of the spectral property for the free Hamiltonian (i.e. Laplace operator) on $S^{1}\setminus\{0\}$.
As a by-product we find three-parameter family of trace formulae which provide a direct connection between quantum energy spectrum and classical length spectrum of $S^{1}$ with a point singularity.
These can be regarded as generalizations of the Poisson summation formula.
In Section \ref{sec:proof} we give a proof of \eqref{eq:1-3}.
In Section \ref{sec:examples} we present explicit examples of the Feynman kernels for several subfamilies of the $U(2)$ family of point interactions.
We conclude in Section \ref{sec:conclusion}.

\section{Reflection and transmission coefficients on $\mathbb{R}\setminus\{0\}$} \label{sec:RT}
Particle collision and production processes are absent from the one-particle quantum mechanics with a single point interaction.
Nevertheless, the reflection and transmission from the point interaction give rise to nontrivial scattering matrix.
In this section we will calculate the matrix elements of scattering matrix, that is, the reflection and transmission coefficients for a continuum state once scattered by the point interaction at the origin on a whole line.

The reflection and transmission coefficients for right- and left-moving incidental waves with momentum $k>0$ are given by
\begin{subequations}
\begin{align}
\psi_{k,+}(x)
&= 	\begin{cases}
	\mathrm{e}^{ikx} + R_{+}(k)\mathrm{e}^{-ikx}, 	& \text{for~} x<0, \\
	T_{+}(k)\mathrm{e}^{ikx}, 				& \text{for~} x>0,
	\end{cases} \label{eq:2-1a}
\end{align}
and
\begin{align}
\psi_{k,-}(x)
&= 	\begin{cases}
	T_{-}(k)\mathrm{e}^{-ikx}, 				& \text{for~} x<0, \\
	\mathrm{e}^{-ikx} + R_{-}(k)\mathrm{e}^{ikx}, 	& \text{for~} x>0.
	\end{cases} \label{eq:2-1b}
\end{align}
\end{subequations}
Point interactions consistent with the probability conservation $j(0_{+}) = j(0_{-})$ are characterized by the same boundary conditions as \eqref{eq:1-8} but with the two-component vectors
\begin{align}
{\vec \Psi}(x)
= 	\begin{pmatrix}
	\psi(x) \\
	\psi(-x)
	\end{pmatrix}, \quad
{\vec \Psi}^{\prime}(x)
= 	\begin{pmatrix}
	\psi^{\prime} \\
	- \psi^{\prime}(-x)
	\end{pmatrix}, \quad
0<x<\infty. \label{eq:2-2}
\end{align}
Plugging \eqref{eq:2-1a} and \eqref{eq:2-1b} into the boundary conditions \eqref{eq:1-8} with \eqref{eq:2-2} we get the matrix equations
\begin{align}
P_{\pm}Z(k)
&= 	\mathrm{e}^{i\delta_{\pm}(k)}P_{\pm}, \label{eq:2-3}
\end{align}
where
\begin{align}
Z(k)
&:= 	\begin{pmatrix}
	R_{-}(k) 	& T_{+}(k) \\
	T_{-}(k) 	& R_{+}(k)
	\end{pmatrix}, \label{eq:2-4}
\end{align}
and
\begin{align}
0\leq
\delta_{\pm}(k)
:= 	2\mathrm{Arccot}(kL_{\pm})
= 	\frac{1}{i}\mathrm{Log}
	\left(\frac{ikL_{\pm} - 1}{ikL_{\pm} + 1}\right)
< 	2\pi. \label{eq:2-5}
\end{align}
$\mathrm{Arccot}$ and $\mathrm{Log}$ are the principal values of the inverse cotangent and logarithm, respectively.
Now it is easy to find the reflection and transmission coefficients.
Equation \eqref{eq:2-3} implies that the matrix $Z(k)$ is unitary and has the spectral decomposition $Z(k) = \mathrm{e}^{i\delta_{+}(k)}P_{+} + \mathrm{e}^{i\delta_{-}(k)}P_{-}$.
Thus,
\begin{align}
Z(k)
&= 	\frac{\mathrm{e}^{i\delta_{+}(k)} + \mathrm{e}^{i\delta_{-}(k)}}{2}\1
	+ \frac{\mathrm{e}^{i\delta_{+}(k)} - \mathrm{e}^{i\delta_{-}(k)}}{2}\vec{e}\cdot\vec{\sigma}, \label{eq:2-6}
\end{align}
from which we find
\begin{subequations}
\begin{align}
R_{\pm}(k)
&= 	\frac{\mathrm{e}^{i\delta_{+}(k)} + \mathrm{e}^{i\delta_{-}(k)}}{2}
	\mp \frac{\mathrm{e}^{i\delta_{+}(k)} - \mathrm{e}^{i\delta_{-}(k)}}{2}e_{z}, \label{eq:2-7a}\\
T_{\pm}(k)
&= 	\frac{\mathrm{e}^{i\delta_{+}(k)} - \mathrm{e}^{i\delta_{-}(k)}}{2}(e_{x} \mp ie_{y}). \label{eq:2-7b}
\end{align}
\end{subequations}
These results are consistent with those obtained in Ref.\cite{Fulop:2001a,Caudrelier:2005} with suitable redefinitions of the parameters.

Several remarks are now in order.
\begin{enumerate}

\item The two eigenvalues of $Z(k)$ satisfy the relations $\mathrm{e}^{i\delta_{\pm}(-k)} = \mathrm{e}^{-i\delta_{\pm}(k)}$, from which we define
\begin{align}
\delta_{\pm}(-k)
= 	\begin{cases}
	2\pi - \delta_{\pm}(k) 	& \text{for}~\alpha_{\pm}\neq0, \\
	- \delta_{\pm}(k) 	& \text{for}~\alpha_{\pm}=0.
	\end{cases} \label{eq:2-8}
\end{align}
Note that this definition ensures the continuity of $\delta_{\pm}(k)$ ($-\infty<k<\infty$) at $k=0$.

\item The \textit{phase shifts} $\delta_{\pm}(k)$ satisfy the following functional identities
\begin{align}
\delta_{\pm}^{\prime}(k)
&= 	- \frac{\sin\delta_{\pm}(k)}{k}, \label{eq:2-9}
\end{align}
where prime ($\prime$) indicates the derivative with respect to $k$.
These identities will be important for the proof of \eqref{eq:1-3}.

\item The reflection and transmission coefficients satisfy
\begin{align}
[R_{\pm}(k)]^{*} = R_{\pm}(-k), \quad
[T_{\pm}(k)]^{*} = T_{\mp}(-k), \label{eq:2-10}
\end{align}
where $\ast$ indicates the complex conjugation.

\item In terms of the reflection and transmission coefficients the unitarity conditions of the matrix $Z(k)$ read
\begin{subequations}
\begin{align}
T_{\mp}(-k)T_{\pm}(k) + R_{\pm}(-k)R_{\pm}(k)
&= 	1, \label{eq:2-11a}\\
T_{\mp}(-k)R_{\mp}(k) + R_{\pm}(-k)T_{\mp}(k)
&= 	0. \label{eq:2-11b}
\end{align}
\end{subequations}

\item The phase shifts $\delta_{\pm}(k)$ become independent of momentum $k$ if $\alpha_{\pm} = 0$ or $\pi$: $\delta_{\pm}(k) = 0$ for $\alpha_{\pm} = 0$ and $\delta_{\pm}(k) = \pi$ for $\alpha_{\pm} = \pi$.
This is important for discussing the scale-independent point interactions in Section \ref{sec:examples}.
\end{enumerate}

\section{Scattering matrix on $\mathbb{R}\setminus\{0\}$} \label{sec:S-matrix}
In this section we will first introduce the one-particle scattering matrix (S-matrix) and then define the $n$-times scattering matrix, whose elements give the weight factors of the Feynman kernel for the contributions scattered $n$-times by the point interaction.

\begin{figure}[t]
	\begin{center}
		\begin{tabular}{lr}
			$
			S^{(1)}_{++}(k)
			= 	\begin{matrix}
				\includegraphics{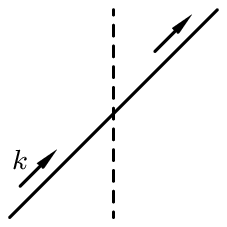}
				\end{matrix}
			$
		&
			$
			\begin{matrix}
			\includegraphics{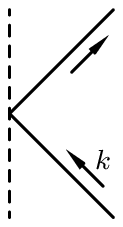}
			\end{matrix}
			= 	S^{(1)}_{+-}(k)
			$
		\\
		& \\
			$
			S^{(1)}_{-+}(k)
			= 	\begin{matrix}
				\includegraphics{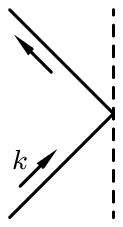}
				\end{matrix}
			$
		&
			$
			\begin{matrix}
			\includegraphics{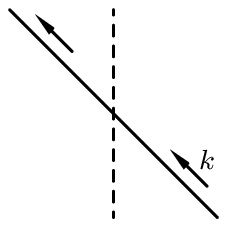}
			\end{matrix}
			= 	S^{(1)}_{--}(k)
			$
		\end{tabular}
		\caption{One-particle scattering from a point interaction (point scatterer). Time is flowing along the vertical direction. The dashed line represents the world line for the point scatterer. The arrows represent the direction of momentum flow.}
		\label{fig:S-matrix}
	\end{center}
\end{figure}

Let us first define the one-particle S-matrix $S^{(1)}$ on a whole line in the presence of a single point interaction at the origin.
In the basis of right- and left-moving momentum mode $\{|+k\rangle, |-k\rangle \mid k>0\}$, where $\langle x|\pm k\rangle = \mathrm{e}^{\pm ikx}$, the one-particle S-matrix is defined as follows:
\begin{align}
S^{(1)}(k)
&= 	\begin{pmatrix}
	S^{(1)}_{++}(k) 	& S^{(1)}_{+-}(k) \\
	S^{(1)}_{-+}(k) 	& S^{(1)}_{--}(k)
	\end{pmatrix}
:= 	\begin{pmatrix}
	T_{+}(k) 	& R_{-}(k) \\
	R_{+}(k) 	& T_{-}(k)
	\end{pmatrix}, \label{eq:3-1}
\end{align}
whose matrix elements are graphically represented in Figure \ref{fig:S-matrix}.
Noting that the S-matrix can be written as $S^{(1)}(k) = Z(k)\sigma_{1}$ and $Z(k)$ is unitary, we see that $S^{(1)}(k)$ clearly satisfies the unitarity conditions $\bigl[S^{(1)}(k)\bigr]^{\dagger}S^{(1)}(k) = \1 = S^{(1)}(k)\bigl[S^{(1)}(k)\bigr]^{\dagger}$, which are nothing but the consequence of the probability conservation $j(0_{+}) = j(0_{-})$.
For the following discussions it is convenient to rewrite the S-matrix into the spectral decomposition form
\begin{align}
S^{(1)}(k)
&= 	s_{+}(k)\mathcal{P}_{+}(k)
	+ s_{-}(k)\mathcal{P}_{-}(k), \label{eq:3-2}
\end{align}
where $s_{\pm}(k)$ are the two eigenvalues of $S^{(1)}(k)$ given by
\begin{subequations}
\begin{align}
s_{\pm}(k)
&= 	\mathrm{e}^{i\left(\Delta_{+}(k) + \pi/2\right)}
	\left[
	e_{x}\sin\Delta_{-}(k)
	\pm i
	\sqrt{1 - e_{x}^{2}\sin^{2}\Delta_{-}(k)}
	\right], \label{eq:3-3a}\\
\Delta_{\pm}(k)
&:= 	\frac{\delta_{+}(k) \pm \delta_{-}(k)}{2}, \label{eq:3-3b}
\end{align}
\end{subequations}
and $\mathcal{P}_{\pm}(k)$ are the corresponding projection operators constructed as follows:
\begin{align}
\mathcal{P}_{\pm}(k)
= 	\frac{S^{(1)}(k) - s_{\mp}(k)\1}{s_{\pm}(k) - s_{\mp}(k)}
= 	\frac{\1 \pm \vec{\varepsilon}(k)\cdot\vec{\sigma}}{2}, \label{eq:3-4}
\end{align}
where $\vec{\varepsilon}(k) = {}^{t}(\varepsilon_{x}(k), \varepsilon_{y}(k), \varepsilon_{z}(k))$ is a real unit vector defined as
\begin{align}
\vec{\varepsilon}(k)
&:= 	\frac{1}{\sqrt{1 - e_{x}^{2}\sin^{2}\Delta_{-}(k)}}
	\begin{pmatrix}
	-\cos\Delta_{-}(k) \\
	e_{z}\sin\Delta_{-}(k) \\
	- e_{y}\sin\Delta_{-}(k)
	\end{pmatrix}. \label{eq:3-5}
\end{align}
Notice that these projection operators $\mathcal{P}_{\pm}(k)$ satisfy the relations $\mathcal{P}_{+}(k) + \mathcal{P}_{-}(k) = \1$, $[\mathcal{P}_{\pm}(k)]^{2} = \mathcal{P}_{\pm}(k)$ and $\mathcal{P}_{\pm}(k)\mathcal{P}_{\mp}(k) = 0$ and that $s_{\pm}(k)$ satisfy the relations
\begin{align}
s_{\pm}(-k)
&= 	\begin{cases}
	1/s_{\pm}(k) 	& \text{for}~\alpha_{\pm}\neq0~\text{or}~\alpha_{\pm}=0, \\
	1/s_{\mp}(k) 	& \text{for}~\alpha_{+}=0, \alpha_{-}\neq0~\text{or}~\alpha_{+}\neq0, \alpha_{-}=0.
	\end{cases} \label{eq:3-6}
\end{align}

Next introduce the $n$-times scattering matrix $S^{(n)}$ as the $n$-th power of $S^{(1)}$:
\begin{align}
S^{(n)}(k)
&= 	\begin{pmatrix}
	S^{(n)}_{++}(k) 	& S^{(n)}_{+-}(k) \\
	S^{(n)}_{-+}(k) 	& S^{(n)}_{--}(k)
	\end{pmatrix}
:= 	\bigl[S^{(1)}(k)\bigr]^{n}. \label{eq:3-7}
\end{align}
By construction it is obvious that the $n$-times scattering matrix $S^{(n)}(k)$ satisfies the unitarity conditions $\bigl[S^{(n)}(k)\bigr]^{\dagger}S^{(n)}(k) = \1 = S^{(n)}(k)\bigl[S^{(n)}(k)\bigr]^{\dagger}$, which lead the partial amplitude unitarity of the Feynman kernel.
Thanks to the spectral decomposition \eqref{eq:3-2} the $n$-times scattering matrix is easily computed with the result
\begin{align}
S^{(n)}(k) = [s_{+}(k)]^{n}\mathcal{P}_{+}(k) + [s_{-}(k)]^{n}\mathcal{P}_{-}(k). \label{eq:3-8}
\end{align}
Although in the following discussions we do not need the explicit expression for $S^{(n)}$, it may be instructive to write down its matrix elements.
A straightforward calculation yields
\begin{subequations}
\begin{align}
S^{(n)}_{\pm\pm}
&= 	\mathrm{e}^{in(\Delta_{+} + \pi/2)}
	\biggl[
	T_{n}\bigl(e_{x}\sin\Delta_{-}\bigr)
	\mp ie_{y}\sin\Delta_{-}\cdot
	U_{n-1}\bigl(e_{x}\sin\Delta_{-}\bigr)
	\biggr], \label{eq:3-9a}\\
S^{(n)}_{\mp\pm}
&= 	\mathrm{e}^{in(\Delta_{+} + \pi/2)}
	\bigl[
	\mp e_{z}\sin\Delta_{-}
	- i\cos\Delta_{-}
	\bigr]
	U_{n-1}\bigl(e_{x}\sin\Delta_{-}\bigr), \label{eq:3-9b}
\end{align}
\end{subequations}
where $T_{n}$ and $U_{n}$ are the Chebyshev polynomials of the first and second kind, respectively, and satisfy the following relations:
\begin{align}
T_{n}(\cos\theta)
&= 	\cos(n\theta), \quad
U_{n}(\cos\theta)
= 	\frac{\sin\bigl((n+1)\theta\bigr)}{\sin\theta}, \quad
n=0,1,2,\cdots. \label{eq:3-10}
\end{align}

\section{Spectrum of $S^{1}\setminus\{0\}$} \label{sec:spectrum}
Let us next study the spectrum of the quantum system for a particle on a circle in the presence of a point interaction described by the boundary conditions \eqref{eq:1-4}.
Although the spectral property of the system has been already studied in the literature \cite{Fulop:1999,Fulop:2003}, those results are not suitable for the purpose of this paper.
In this Section we will uncover an amazing relation between the scattering theory on $\mathbb{R}\setminus\{0\}$ discussed in the previous section and positive energy spectrum of $S^{1}\setminus\{0\}$.
We also derive the trace formulae for the free Hamiltonian (Laplace operator) on $S^{1}\setminus\{0\}$.

The general solution to the Schr\"odinger equation $-\mathrm{d}^{2}\psi/\mathrm{d}x^{2} = E\psi$ on $S^{1}\setminus\{0\}$ for positive energy $E = k^{2} > 0$ is given by
\begin{align}
\psi_{k}(x)
&= 	A(k)\mathrm{e}^{ikx} + B(k)\mathrm{e}^{ik(L-x)}, \quad
k>0, \label{eq:4-1}
\end{align}
where the phase factor $\mathrm{e}^{ikL}$ in the second term is introduced for the later convenience.
Notice that the two coefficients $A(k)$ and $B(k)$ may depend on $k$.
The general solution for negative energy $E=-\kappa^{2}<0$ will be obtained by just replacing $k$ to $i\kappa$ in \eqref{eq:4-1}.
We have to be, however, careful about zero energy solutions with $E=0$, which are \textit{not} necessarily obtained by the naive limit $k\to0$ in \eqref{eq:4-1}: The general solution for $E=0$ is a first degree polynomial and takes the form
\begin{align}
\psi_{0}(x)
&= 	A_{0} + B_{0}x.
\end{align}
In this paper, we call the above solution with $B_{0} \neq 0$, as well as negative energy solutions, \textit{bound states}.
It turns out that any knowledge about those bound states are not necessary for the following discussions.
Notice that a zero energy solution with the limit $k\to0$ in \eqref{eq:4-1} is ambiguous because the two terms in \eqref{eq:4-1} are not independent each other when $k=0$.
This issue will be discussed later.

Substituting \eqref{eq:4-1} into \eqref{eq:1-8} we get the two independent conditions
\begin{align}
P_{\pm}
\left(
\mathrm{e}^{-ikL}\1 - \mathrm{e}^{i\delta_{\pm}(k)}\sigma_{1}
\right)
\begin{pmatrix}
A(k) \\
B(k)
\end{pmatrix}
&= 	\vec{0}. \label{eq:4-2}
\end{align}
Since these two equations are orthogonal to each other, they can be combined into the following form:
\begin{align}
S^{(1)}(k)
\begin{pmatrix}
A(k) \\
B(k)
\end{pmatrix}
&= 	\mathrm{e}^{-ikL}
	\begin{pmatrix}
	A(k) \\
	B(k)
	\end{pmatrix}, \label{eq:4-3}
\end{align}
which follows from $P_{+} + P_{-} = \1$ and $S^{(1)} = (\mathrm{e}^{i\delta_{+}}P_{+} + \mathrm{e}^{i\delta_{-}}P_{-})\sigma_{1}$.
This eigenvalue equation indicates that the positive energy spectrum and eigenfunctions of single particle quantum mechanics on $S^{1}\setminus\{0\}$ is completely determined by the one-particle S-matrix on $\mathbb{R}\setminus\{0\}$.

In the following we will analyze the eigenvalue equation \eqref{eq:4-3} in detail.

\subsection{Spectrum quantization conditions} \label{sec:quantization}
Let us first study the spectral property of $S^{1}\setminus\{0\}$.
For non-vanishing $A(k)$ and $B(k)$ we have to implement the following condition
\begin{align}
\det
\left[
S^{(1)}(k) - \mathrm{e}^{-ikL}\1
\right]
&= 	0, \quad
k>0, \label{eq:4-1-1}
\end{align}
which has two branches
\begin{align}
\mathrm{e}^{-ikL}
&= 	s_{+}(k) \quad\text{and}\quad
\mathrm{e}^{-ikL}
= 	s_{-}(k), \label{eq:4-1-2}
\end{align}
where $s_{\pm}(k)$ are given in \eqref{eq:3-3a}.
It should be pointed out that these types of equations are commonly referred to as the Bethe ansatz equations.
Indeed, the generalization to $n$-particle system has been studied in the literature \cite{Caudrelier:2005} under the name of impurity Bethe ansatz equation.

The positive energy spectrum is determined as the positive roots of the equations
\begin{align}
f_{\pm}(k)
&= 	0, \label{eq:4-1-3}
\end{align}
where
\begin{align}
f_{\pm}(k)
:= 	kL + \frac{1}{i}\log s_{\pm}(k). \label{eq:4-1-4}
\end{align}
It should be emphasized that we thought the logarithm function as the multivalued function defined as $\log z = \{\ln |z| + i\mathrm{Arg}z + i2m\pi \mid 0\leq\mathrm{Arg}z <2\pi, m\in\mathbb{Z}\}$, where $\mathrm{Arg}z$ is the principal value of the argument.
Each integer $m$ determines the branch of the logarithm function and $m=0$ corresponds to the principal branch.

Note that if $\lim_{k\to0}s_{\pm}(k) = 1$, the equation \eqref{eq:4-1-3} may have zero energy solutions.
However, the existence of such a zero energy solution dose not necessarily imply a physical state in the spectrum because the $k=0$ solution in \eqref{eq:4-1} becomes trivial and should be thrown away if $A(0) + B(0) = 0$, even though both $A(0)$ and $B(0)$ are not identically zero.
Nevertheless, it turns out that such a fake solution is necessary in the trace formulae discussed in the next subsection and the proof of \eqref{eq:1-3}.

\subsection{Trace formulae} \label{sec:trace_formulae}
In order to fulfill the gap between the operator formalism and the path-integral formalism, we have to establish the trace formulae for $S^{1}\setminus\{0\}$.
To this end, let us consider delta functions $\delta\bigl(f_{\pm}(k)\bigr)$.
Since the values assumed by $f_{\pm}(k)$ are $kL + \mathrm{Arg}[s_{\pm}(k)] + 2m\pi$ for all integers $m$, the delta functions $\delta\bigl(f_{\pm}(k)\bigr)$ are periodic functions of $f_{\pm}$ with a period $2\pi$ so that it can be expanded into the Fourier series
\begin{align}
\delta\bigl(f_{\pm}(k)\bigr)
&= 	\frac{1}{2\pi}\sum_{n\in\mathbb{Z}}\mathrm{e}^{inf_{\pm}(k)}. \label{eq:4-2-1}
\end{align}
Note that the left hand side can be written as $\sum_{k_{m}^{\pm}\in\sigma_{\pm}}(1/\bigl|f_{\pm}^{\prime}(k_{m}^{\pm})\bigr|)\delta(k - k_{m}^{\pm})$, where $\sigma_{\pm}$ are the sets of both positive and negative roots of the equations $f_{\pm}(k) = 0$ defined as
\begin{align}
\sigma_{\pm}
:= 	\bigl\{k_{m}^{\pm}\in\mathbb{R}
	\mid f_{\pm}(k_{m}^{\pm}) = 0,
	\cdots<k_{-2}^{\pm}<k_{-1}^{\pm}<k_{0}^{\pm}=0<k_{1}^{\pm}<k_{2}^{\pm}<\cdots\bigr\}. \label{eq:4-2-2}
\end{align}
Eq.\eqref{eq:4-2-2} requires more detail explanations:
\begin{itemize}
\item The negative roots ($m<0$) are related to the positive ones as follows:
	\begin{align}
	k_{-m}^{\pm}
	&= 	\begin{cases}
		- k_{m}^{\pm} 	& \text{for}~\alpha_{\pm}\neq0~\text{or}~\alpha_{\pm}=0, \\
		- k_{m}^{\mp} 	& \text{for}~\alpha_{+} = 0, \alpha_{-} \neq 0~\text{or}~\alpha_{+} \neq 0, \alpha_{-} = 0,
		\end{cases} \label{eq:4-2-4}
	\end{align}
	which follow from \eqref{eq:3-6} and \eqref{eq:4-1-2}.
	These relations will be used in the proof of \eqref{eq:1-3}.
\item The $k_{0}^{\pm} = 0$ roots appear only in the following four cases:
	\begin{subequations}
	\begin{align}
	k_{0}^{+} 		& \quad\text{for}~\alpha_{\pm}\neq0, \label{eq:4-2-3a}\\
	k_{0}^{\pm} 	& \quad\text{for}~\alpha_{+}=0, \alpha_{-}\neq0, e_{x}=1, \label{eq:4-2-3b}\\
	k_{0}^{\pm} 	& \quad\text{for}~\alpha_{+}\neq0, \alpha_{-}=0, e_{x}=-1, \label{eq:4-2-3c}\\
	k_{0}^{-} 		& \quad\text{for}~\alpha_{\pm}=0. \label{eq:4-2-3d}
	\end{align}
	\end{subequations}
	It turns out that the solution of $k_{0}^{+} = 0$ for $\alpha_{\pm}\neq0$ and one of the two $k_{0}^{\pm} = 0$ solutions for $\alpha_{+} = 0$, $\alpha_{-}\neq0$, $e_{x}=1$ or $\alpha_{+}\neq-$, $\alpha_{-} = 0$, $e_{x}=-1$ are fake solutions with $A(0) + B(0) = 0$, as explained in the previous subsection.
	It is emphasized that the $k_{0}^{\pm} = 0$ solutions (if they exist) must be included in $\sigma_{\pm}$, irrespective of a fake or genuine zero mode.
\end{itemize}
We note that these remarks will be important for the proof of \eqref{eq:1-3}, however, they are not relevant for the rest of this subsection.

Now, the identity \eqref{eq:4-2-1} becomes the following 3-parameter family of the trace formulae
\begin{align}
\sum_{k_{m}^{\pm}\in\sigma_{\pm}}
\frac{1}{|f_{\pm}^{\prime}(k_{m}^{\pm})|}
\delta(k - k_{m}^{\pm})
&= 	\frac{1}{2\pi}
	\sum_{n\in\mathbb{Z}}\mathrm{e}^{inf_{\pm}(k)}, \label{eq:4-2-5}
\end{align}
which include the Poisson summation formula as a certain region of the parameter space spanned by $\alpha_{+}$, $\alpha_{-}$ and $e_{x}$.
Notice that the derivatives of $f_{\pm}(k)$ are given as follows:
\begin{align}
f_{\pm}^{\prime}(k)
=	L + \frac{\delta_{+}^{\prime}(k) + \delta_{-}^{\prime}(k)}{2}
	\pm \frac{\delta_{+}^{\prime}(k) - \delta_{-}^{\prime}(k)}{2}
	\frac{-e_{x}\cos\Delta_{-}(k)}
	{\sqrt{1 - e_{x}^{2}\sin^{2}\Delta_{-}(k)}}, \label{eq:4-2-6}
\end{align}
and satisfy
\begin{align}
f_{\pm}^{\prime}(-k)
&= 	\begin{cases}
	f_{\pm}^{\prime}(k) 	& \text{for}~\alpha_{\pm}\neq0~\text{or}~\alpha_{\pm}=0, \\
	f_{\mp}^{\prime}(k) 	& \text{for}~\alpha_{+}=0, \alpha_{-}\neq0~\text{or}~\alpha_{+}\neq0, \alpha_{-}=0.
	\end{cases} \label{eq:4-2-7}
\end{align}
As we will see in Section \ref{sec:eigenfunction}, $f_{\pm}^{\prime}(k_{m}^{\pm})$ give the normalization factors for the positive energy eigenfunctions.

Before closing this subsection we try to rewrite the formulae \eqref{eq:4-2-5} in more practically convenient expression.
Since $\sum_{k_{m}^{\pm}\in\sigma_{\pm}}(1/|f_{\pm}^{\prime}(k_{m}^{\pm})|)\delta(k - k_{m}^{\pm}) = (1/|f_{\pm}^{\prime}(k)|)\sum_{k_{m}^{\pm}\in\sigma_{\pm}}\delta(k - k_{m}^{\pm})$, \eqref{eq:4-2-5} can be rewritten into the form $\sum_{k_{m}^{\pm}\in\sigma_{\pm}}\delta(k - k_{m}^{\pm}) = (1/2\pi)|f_{\pm}^{\prime}(k)|\sum_{n\in\mathbb{Z}}\mathrm{e}^{inf_{\pm}(k)}$.
Thus, by multiplying a smooth test function $F(k)$ and integrating out over the range $-\infty<k<\infty$, the trace formulae \eqref{eq:4-2-5} can be cast into the following form:
\begin{align}
\sum_{k_{m}^{\pm}\in\sigma_{\pm}} F(k_{m}^{\pm})
&= 	\sum_{n\in\mathbb{Z}}
	\int_{-\infty}^{\infty}\!\frac{\mathrm{d}k}{2\pi}
	\left|\frac{\mathrm{d}f_{\pm}(k)}{\mathrm{d}k}\right|
	F(k)\mathrm{e}^{inf_{\pm}(k)}. \label{eq:4-2-8}
\end{align}
This identity will be useful for computations of the Casimir energy or the perturbative loop calculations of Feynman diagrams in quantum field theory with nontrivial extended defects (branes or boundaries).

\subsection{Reconstruction of S-matrix} \label{sec:reconstruction}
Once given the two eigenvalues $s_{\pm}(k)$ and the corresponding complete orthonormal eigenvectors $|\pm\rangle := {}^{t}(A^{\pm}(k), B^{\pm}(k))$ satisfying $\langle \pm|\pm\rangle = 1$, $\langle \pm|\mp\rangle = 0$ and $|+\rangle\langle +| + |-\rangle\langle -| = \1$, the one-particle S-matrix can be reconstructed in terms of $A^{\pm}(k)$ and $B^{\pm}(k)$ from the projection operators
\begin{align}
\mathcal{P}_{\pm}(k)
&= 	|\pm\rangle\langle\pm|
=	\begin{pmatrix}
	|A^{\pm}(k)|^{2} 			& A^{\pm}(k)[B^{\pm}(k)]^{\ast} \\
	[A^{\pm}(k)]^{\ast}B^{\pm}(k) 	& |B^{\pm}(k)|^{2}
	\end{pmatrix}, \label{eq:4-3-1}
\end{align}
which of course satisfy $\mathcal{P}_{+}(k) + \mathcal{P}_{-}(k) = \1$, $[\mathcal{P}_{\pm}(k)]^{2} = \mathcal{P}_{\pm}(k)$ and $\mathcal{P}_{\pm}(k)\mathcal{P}_{\mp}(k) = 0$.
It follows from the orthonormality and completeness that
\begin{subequations}
\begin{align}
& 	|A^{+}(k)|^{2} + |A^{-}(k)|^{2} = 1, \label{eq:4-3-2a}\\
& 	|B^{+}(k)|^{2} + |B^{-}(k)|^{2} = 1, \label{eq:4-3-2b}\\
& 	A^{+}(k)[B^{+}(k)]^{\ast} + A^{-}(k)[B^{-}(k)]^{\ast} = 0, \label{eq:4-3-2c}\\
& 	[A^{+}(k)]^{\ast}B^{+}(k) + [A^{-}(k)]^{\ast}B^{-}(k) = 0. \label{eq:4-3-2d}
\end{align}
\end{subequations}
By comparing \eqref{eq:4-3-1} to \eqref{eq:3-4} we find the following relations
\begin{subequations}
\begin{align}
|B^{\pm}(-k)|^{2}
&= 	\begin{cases}
	|A^{\pm}(k)|^{2} 	& \text{for}~\alpha_{\pm}\neq0~\text{or}~\alpha_{\pm}=0, \\
	|A^{\mp}(k)|^{2} 	& \text{for}~\alpha_{+}=0, \alpha_{-}\neq0~\text{or}~\alpha_{+}\neq0, \alpha_{-}=0,
	\end{cases} \label{eq:4-3-3a}\\
A^{\pm}(-k)[B^{\pm}(-k)]^{\ast}
&= 	\begin{cases}
	[A^{\pm}(k)]^{\ast}B^{\pm}(k) 	& \text{for}~\alpha_{\pm}\neq0~\text{or}~\alpha_{\pm}=0, \\
	[A^{\mp}(k)]^{\ast}B^{\mp}(k) 	& \text{for}~\alpha_{+}=0, \alpha_{-}\neq0~\text{or}~\alpha_{+}\neq0, \alpha_{-}=0.
	\end{cases} \label{eq:4-3-3b}
\end{align}
\end{subequations}
Each component of the $n$-times scattering matrix $S^{(n)}(k)$ in \eqref{eq:3-6} is found to be
\begin{subequations}
\begin{align}
S_{++}^{(n)}(k)
&= 	\sum_{\xi=\pm}[s_{\xi}(k)]^{n}|A^{\xi}(k)|^{2}, \label{eq:4-3-4a}\\
S_{--}^{(n)}(k)
&= 	\sum_{\xi=\pm}[s_{\xi}(k)]^{n}|B^{\xi}(k)|^{2}, \label{eq:4-3-4b}\\
S_{-+}^{(n)}(k)
&= 	\sum_{\xi=\pm}[s_{\xi}(k)]^{n}[A^{\xi}(k)]^{\ast}B^{\xi}(k), \label{eq:4-3-4c}\\
S_{+-}^{(n)}(k)
&= 	\sum_{\xi=\pm}[s_{\xi}(k)]^{n}A^{\xi}(k)[B^{\xi}(k)]^{\ast}. \label{eq:4-3-4d}
\end{align}
\end{subequations}

\subsection{Eigenfunctions} \label{sec:eigenfunction}
In terms of the orthonormal eigenvectors $|\pm\rangle = {}^{t}(A^{\pm}(k), B^{\pm}(k))$, the energy eigenfunctions \eqref{eq:4-1} are rewritten as follows:
\begin{align}
\psi_{m}^{\pm}(x)
&= 	N_{m}^{\pm}
	\left[
	A^{\pm}(k_{m}^{\pm})\mathrm{e}^{ik_{m}^{\pm}x}
	+ B^{\pm}(k_{m}^{\pm})\mathrm{e}^{ik_{m}^{\pm}(L - x)}
	\right], \label{eq:4-4-1}
\end{align}
where the normalization factors $N_{m}^{\pm}$ are given by
\begin{align}
|N_{m}^{\pm}|^{2}
=	\left(
	L
	+ \bigl\{A^{\pm}(k_{m}^{\pm})[B^{\pm}(k_{m}^{\pm})]^{\ast}
	+ B^{\pm}(k_{m}^{\pm})[A^{\pm}(k_{m}^{\pm})]^{\ast}\bigr\}
	\frac{\mathrm{e}^{ik_{m}^{\pm}L} - \mathrm{e}^{-ik_{m}^{\pm}L}}{2ik_{m}^{\pm}}
	\right)^{-1}. \label{eq:4-4-2}
\end{align}
With the help of the identities $A^{\pm}(k_{m}^{\pm})[B^{\pm}(k_{m}^{\pm})]^{\ast}+ B^{\pm}(k_{m}^{\pm})[A^{\pm}(k_{m}^{\pm})]^{\ast} = \mathrm{tr}[\mathcal{P}_{\pm}(k_{m}^{\pm})\sigma_{1}] = \pm\varepsilon_{x}(k_{m}^{\pm})$, $\mathrm{e}^{ik_{m}^{\pm}L} - \mathrm{e}^{-ik_{m}^{\pm}L} = -2i\mathrm{Im}\bigl[s_{\pm}(k_{m}^{\pm})\bigr]$ and \eqref{eq:2-9}, it is not difficult to show that the normalization factors can be written as follows
\begin{align}
|N_{m}^{\pm}|^{2}
= 	\frac{1}{f_{\pm}^{\prime}(k_{m}^{\pm})}. \label{eq:4-4-3}
\end{align}

\section{Feynman kernel} \label{sec:proof}
In this Section, we prove our main goal of the formula \eqref{eq:1-3}.
To this end, let us first discuss the case of $0<\alpha_{\pm}<2\pi$.
In the operator formalism the Feynman kernel is then given by
\begin{align}
K(x, T; x_{0}, 0)
&= 	\sum_{\xi=\pm}\sum_{m=1}^{\infty}
	\mathrm{e}^{-i(k_{m}^{\xi})^{2}T}\psi_{m}^{\xi}(x)[\psi_{m}^{\xi}(x_{0})]^{\ast} \nonumber\\
& 	+ (\text{bound state contribution}). \label{eq:5-1}
\end{align}
We note that a fake $k_{0}^{+} = 0$ mode is not included in the above summation, as it should be.
Substituting \eqref{eq:4-4-1} into \eqref{eq:5-1}, we have
\begin{align}
K(x,T; x_{0},0)
&= 	\sum_{\xi=\pm}\sum_{m=1}^{\infty}
	\mathrm{e}^{-i(k_{m}^{\xi})^{2}T}|N_{m}^{\xi}|^{2}
	\Bigl\{
	|A^{\xi}(k_{m}^{\xi})|^{2}\mathrm{e}^{ik_{m}^{\xi}(x-x_{0})}
	+ |B^{\xi}(k_{m}^{\xi})|^{2}\mathrm{e}^{-ik_{m}^{\xi}(x-x_{0})} \nonumber\\
& 	+ [A^{\xi}(k_{m}^{\xi})]^{\ast}B^{\xi}(k_{m}^{\xi})\mathrm{e}^{ik_{m}^{\xi}(L-x-x_{0})}
	+ A^{\xi}(k_{m}^{\xi})[B^{\xi}(k_{m}^{\xi})]^{\ast}\mathrm{e}^{-ik_{m}^{\xi}(L-x-x_{0})}
	\Bigr\} \nonumber\\
& 	+ (\text{bound state contribution}). \label{eq:5-2}
\end{align}
By use of the relations \eqref{eq:4-2-4} \eqref{eq:4-2-7} \eqref{eq:4-3-3a} \eqref{eq:4-3-3b} \eqref{eq:4-4-3} for $0<\alpha_{\pm}<2\pi$, we can rewrite \eqref{eq:5-2} as
\begin{align}
K(x,T; x_{0},0)
&= 	\sum_{\xi=\pm}\sum_{k_{m}^{\xi}\in\sigma_{\xi}}
	\mathrm{e}^{-i(k_{m}^{\xi})^{2}T}\frac{1}{f_{\xi}^{\prime}(k_{m}^{\xi})} \nonumber\\
&	\times\left\{
	|A^{\xi}(k_{m}^{\xi})|^{2}\mathrm{e}^{ik_{m}^{\xi}(x-x_{0})}
 	+ [A^{\xi}(k_{m}^{\xi})]^{\ast}B^{\xi}(k_{m}^{\xi})\mathrm{e}^{ik_{m}^{\xi}(L-x-x_{0})}
	\right\} \nonumber\\
& 	+ (\text{bound state contribution}). \label{eq:5-3}
\end{align}
We should notice that the summations over $k_{m}^{\pm}$ can be enlarged to $\sigma_{\pm}$ and a fake $k_{0}^{+}=0$ mode is added in \eqref{eq:5-3} with the relation $A^{+}(k_{0}^{+}) + B^{+}(k_{0}^{+}) = 0$.
Now we can use the trace formula \eqref{eq:4-2-8}:
\begin{align}
K(x,T; x_{0},0)
&= 	\sum_{\xi=\pm}\sum_{n\in\mathbb{Z}}
	\int_{-\infty}^{\infty}\frac{\mathrm{d}p}{2\pi}[s_{\xi}(p)]^{n} \nonumber\\
&	\times\left\{
	|A^{\xi}(p)|^{2}\mathrm{e}^{iT\left[p\left(\frac{nL+x-x_{0}}{T}\right) - p^{2}\right]}
 	+ [A^{\xi}(p)]^{\ast}B^{\xi}(p)\mathrm{e}^{iT\left[p\left(\frac{(n+1)L-x-x_{0}}{T}\right) - p^{2}\right]}
	\right\} \nonumber\\
& 	+ (\text{bound state contribution}), \label{eq:5-4}
\end{align}
where the definitions \eqref{eq:4-1-4} have been used.
By use of the relations \eqref{eq:3-6} \eqref{eq:4-3-2a} \eqref{eq:4-3-3a} \eqref{eq:4-3-3b} \eqref{eq:4-3-4a}--\eqref{eq:4-3-4d} we finally arrive at the conclusion \eqref{eq:1-3}.

It is interesting to point out that the final expression \eqref{eq:1-3} holds for other values of $\alpha_{\pm}$, even though the relations \eqref{eq:4-2-4} \eqref{eq:4-2-7} \eqref{eq:4-3-3a} \eqref{eq:4-3-3b} and the existence/nonexistence of a fake zero mode, as well as physical zero energy states, depend on $\alpha_{\pm}$.
Furthermore, we emphasize that the knowledge of the energy eigenvalues and eigenstates is required in the expression of the Feynman kernel \eqref{eq:5-1}, while only the one-particle scattering matrix is sufficient to represent the Feynman kernel in our formulation.
This suggests that the expression \eqref{eq:1-3} is more fundamental than the original one \eqref{eq:5-1}.

\section{Examples} \label{sec:examples}
In this Section we present explicit examples of the Feynman kernels for several subfamilies of point interactions, which are partly classified in the literature \cite{Fulop:1999,Fulop:2003} and summarized in Table \ref{tab:2}.
Since the time-reversal invariant subfamily and the $\mathscr{PT}$-symmetric subfamily are quite involved, in what follows we will only consider the cases of the reflectionless subfamily (also known as the smooth subfamily), the scale-independent subfamily, the pure reflection subfamily (also known as the separated subfamily) and the parity invariant subfamily.
As we will see below, our formalism recover all known results.
\begin{table}[t]
\newcommand{\lw}[1]{\smash{\lower2.0ex\hbox{#1}}}
\begin{center}
\caption{Several subfamilies of point interactions \cite{Fulop:1999,Fulop:2003}.}
\label{tab:2}
\renewcommand{\arraystretch}{1.3}
\begin{tabular}{l || l | c | c}
\noalign{\hrule height 1pt}
\lw{subfamily of point interactions} 		& \lw{constraints on $U$} 			& \multicolumn{2}{c}{constraints on parameter space} \\
\cline{3-4}
								& 								& $(\alpha_{+}, \alpha_{-})$ 			& $(e_{x}, e_{y}, e_{z})$ \\
\noalign{\hrule height 1pt}
reflectionless subfamily 				& $\det(U\pm\1) = 0$				& \lw{$(0,\pi)$, $(\pi,0)$}			& \lw{$(e_{x}, e_{y}, 0)$} \\
(smooth subfamily)					& \& $U = \sigma_{1}{}^{t}U\sigma_{1}$ & 								& \\
\hline
\lw{scale-independent subfamily} 		& $U = \pm\1$ 					& $(0,0)$, $(\pi,\pi)$ 				& -- \\
								& $\det(U\pm\1) = 0$ 				& $(0,\pi)$, $(\pi,0)$ 				& $(e_{x}, e_{y}, e_{z})$ \\
\hline
pure reflection subfamily				& \lw{$U = \sigma_{3}U\sigma_{3}$} 	& \lw{$(\alpha_{+}, \alpha_{-})$} 		& \lw{$(0, 0, \pm1)$} \\
(separated subfamily) 				& 								& 								& \\
\hline
$\mathscr{P}$-symmetric subfamily 	& $U = \sigma_{1}U\sigma_{1}$ 		& $(\alpha_{+}, \alpha_{-})$ 			& $(\pm1, 0, 0)$ \\
\hline
$\mathscr{T}$-symmetric subfamily 		& $U = {}^{t}U$ 						& $(\alpha_{+}, \alpha_{-})$ 			& $(e_{x}, 0, e_{z})$ \\
\hline
$\mathscr{PT}$-symmetric subfamily 	& $U = \sigma_{1}{}^{t}U\sigma_{1}$ 	& $(\alpha_{+}, \alpha_{-})$ 			& $(e_{x}, e_{y}, 0)$ \\
\noalign{\hrule height 1pt}
\end{tabular}
\end{center}
\end{table}

\subsection{Reflectionless subfamily} \label{subsec:reflectionless}
Let us first consider the reflectionless point interaction as the simplest example because in this case the S-matrix becomes diagonal.
Since $S^{(1)}(k) = Z(k)\sigma_{1}$, the diagonal S-matrix is obtained by the off-diagonal $Z(k)$, which is given by $e_{z} = 0$ and $(\alpha_{+}, \alpha_{-}) = (0, \pi)$ or $(\pi, 0)$.
Since the difference between the two cases $(\alpha_{+}, \alpha_{-}) = (0, \pi)$ and $(\pi, 0)$ is just the overall sign of the S-matrix, without any loss of generality we can restrict ourselves to the case $(\alpha_{+}, \alpha_{-}) = (0, \pi)$ by using the parameterization $\vec{e} = (\cos\theta, \sin\theta, 0)$, $0\leq\theta<2\pi$.
With this parameterization the transmission coefficients are $T_{\pm} = \mathrm{e}^{\mp i\theta}$ so that the $n$-times scattering matrix becomes
\begin{align}
S^{(n)}(k)
&= 	\begin{pmatrix}
	\mathrm{e}^{-in\theta} 	& 0 \\
	0 					& \mathrm{e}^{in\theta}
	\end{pmatrix}. \label{eq:6-1-1}
\end{align}
The Feynman kernel \eqref{eq:1-3} is cast into the well-known form \cite{Schulman:1981}
\begin{align}
K(x,T; x_{0},0)
&= 	\sum_{n\in\mathbb{Z}}\mathrm{e}^{-in\theta}\int_{-\infty}^{\infty}\frac{\mathrm{d}p}{2\pi}
	\mathrm{e}^{iT\left[p\left(\frac{nL + x - x_{0}}{T}\right) - p^{2}\right]}. \label{eq:6-1-2}
\end{align}
(Notice that there is no negative energy state contribution because the S-matrix has no pole.)

It should be mentioned here the connection between the Laidlaw-DeWitt theorem \cite{Laidlaw:1971} and our formalism.
The theorem states that the path-integral in a multiply-connected space has to be taken the sum over the homotopy class of the space with a weight factor which forms a scalar unitary representation of the fundamental group of the space.
In this well-known example, the weight factor is just a phase $\mathrm{e}^{-in\theta}$.
In the viewpoint of the Laidlaw-DeWitt theorem, this weight factor is a scalar unitary representation of $\pi_{1}(S^{1})\cong\mathbb{Z}$, while in our viewpoint it is just the element of $S^{(n)}$, which is essentially unitary thanks to the unitarity of the S-matrix.

\subsection{Scale-independent subfamily} \label{subsec:scale_invariant}
As a next example let us consider the scale-independent point interactions, which include the above reflectionless point interaction and the $\delta^{\prime}$-interaction discussed in Section \ref{sec:intro}.
It is known that the scale-independent point interactions are characterized by the matrices $U$ whose two eigenvalues are  $(1, 1)$, $(-1, -1)$, $(1, -1)$, $(-1, 1)$: that is, $(\alpha_{+}, \alpha_{-}) = (0, 0)$, $(\pi, \pi)$, $(0, \pi)$, $(\pi, 0)$ in \eqref{eq:1-5} \cite{Fulop:1999,Fulop:2003}.
The first two cases also belong to the purely reflecting point interactions and will be considered in the next example.
The latter two cases, however, are different only for the overall sign of the S-matrix, without any loss of generality we can restrict ourselves to the case $(\alpha_{+}, \alpha_{-}) = (0, \pi)$ by using the parameterization $\vec{e} = (\cos\theta, \sin\theta\cos\phi, \sin\theta\sin\phi)$ with $0\leq\theta\leq\pi$ and $0\leq\phi<2\pi$.
With this choice of the parameters, the S-matrix has the form $S^{(1)} = (\vec{e}\cdot\vec{\sigma})\sigma_{1}$, which is just the constant matrix thanks to the scale-independence of the boundary conditions.

The $n$-the power of the S-matrix is easily computed via \eqref{eq:3-7} with $s_{\pm} =  \mathrm{e}^{\mp i\theta}$ and $\mathcal{P}_{\pm} = (\1\mp\sin\phi\sigma_{2}\pm\cos\phi\sigma_{3})/2$.
Then the $n$-times scattering matrix is cast into the following form
\begin{align}
S^{(n)}(k)
&= 	\begin{pmatrix}
	\cos(n\theta) - i\cos\phi\sin(n\theta) 	& \sin\phi\sin(n\theta) \\
	-\sin\phi\sin(n\theta) 				& \cos(n\theta) + i\cos\phi\sin(n\theta)
	\end{pmatrix}. \label{eq:6-2-1}
\end{align}
Thus the Feynman kernel \eqref{eq:1-3} becomes
\begin{align}
K(x,T; x_{0},0)
&= 	\sum_{n\in\mathbb{Z}}[\cos(n\theta) - i\cos\phi\sin(n\theta)]
	\int_{-\infty}^{\infty}\frac{\mathrm{d}p}{2\pi}
	\mathrm{e}^{iT\left[p\left(\frac{nL + x - x_{0}}{T}\right) - p^{2}\right]} \nonumber\\
& 	-\sum_{n\in\mathbb{Z}}\sin\phi\sin(n\theta)
	\int_{-\infty}^{\infty}\frac{\mathrm{d}p}{2\pi}
	\mathrm{e}^{iT\left[p\left(\frac{(n+1)L - x - x_{0}}{T}\right) - p^{2}\right]}. \label{eq:6-2-2}
\end{align}
Observe that $\phi = 0$ ($\phi = \pi$) reduces to the previous results \eqref{eq:6-1-2} for $0\leq\theta\leq\pi$ ($\pi\leq\theta<2\pi$).
The case $\phi = \pi/2$ ($\phi = 3\pi/2$) becomes the kernel \eqref{eq:1-1} for $c>0$ ($c<0$).

The kernel \eqref{eq:6-2-2} coincides with the results evaluated in the operator formalism in \cite{Fulop:1999,Fulop:2003} with suitable redefinitions of the parameters.

\subsection{Pure reflection subfamily} \label{subsec:pure_reflection}
Next consider the purely reflecting point interactions.
These point interactions are obtained from the diagonal $Z(k)$, which is realized by $e_{x} = e_{y} = 0$.
With this choice the transmission coefficients identically vanish and the reflection coefficients become $R_{\pm}(k) = \mathrm{e}^{i\delta_{\mp}(k)}$ for $e_{z} = 1$ and $R_{\pm}(k) = \mathrm{e}^{i\delta_{\pm}(k)}$ for $e_{z} = -1$.
In the following we will consider the case $e_{z} = 1$.
The result for $e_{z} = -1$ will be obtained by replacing $\delta_{+}$ to $\delta_{-}$ and vice versa.
In the $e_{z} = 1$ case the $n$-times scattering matrix $S^{(n)}$ has the form
\begin{align}
S^{(n)}(k)
&= 	\begin{cases}
	\begin{pmatrix}
	\mathrm{e}^{i\frac{n}{2}\delta_{+}(k)}\mathrm{e}^{i\frac{n}{2}\delta_{-}(k)} 	& 0 \\
	0 															& \mathrm{e}^{i\frac{n}{2}\delta_{+}(k)}\mathrm{e}^{i\frac{n}{2}\delta_{-}(k)}
	\end{pmatrix},
	& \text{for}~n=2, 4, 6, \cdots, \\
	\begin{pmatrix}
	0 										& \mathrm{e}^{i\frac{n+1}{2}\delta_{+}(k)}\mathrm{e}^{i\frac{n-1}{2}\delta_{-}(k)} \\
	\mathrm{e}^{i\frac{n-1}{2}\delta_{+}(k)}\mathrm{e}^{i\frac{n+1}{2}\delta_{-}(k)} 	& 0
	\end{pmatrix},
	& \text{for}~n=1, 3, 5, \cdots.
	\end{cases} \label{eq:6-3-1}
\end{align}
In this case of the pure reflection subfamily, the physical meanings of $\delta_{+}$ and $\delta_{-}$ are clear:
$\delta_{-}$ ($\delta_{+}$) is nothing but the \textit{phase shift} which occurs every time when a particle hits the reflecting wall at $x=L$ ($x=0$) from the left (right).
This is the reason why we call these $\delta_{\pm}$ the phase shifts.

Noting that the relations \eqref{eq:2-8}, we can rewrite the kernel \eqref{eq:1-3} into the following form
\begin{align}
K(x,T; x_{0},0)
&= 	\sum_{n\in\mathbb{Z}}\int_{-\infty}^{\infty}\frac{\mathrm{d}p}{2\pi}
	\mathrm{e}^{in\delta_{+}(p)}\mathrm{e}^{in\delta_{-}(p)}
	\mathrm{e}^{iT\left[p\left(\frac{2nL + x - x_{0}}{T}\right) - p^{2}\right]} \nonumber\\
&	+
	\sum_{n\in\mathbb{Z}}\int_{-\infty}^{\infty}\frac{\mathrm{d}p}{2\pi}
	\mathrm{e}^{i(n-1)\delta_{+}(p)}\mathrm{e}^{in\delta_{-}(p)}
	\mathrm{e}^{iT\left[p\left(\frac{2nL - x - x_{0}}{T}\right) - p^{2}\right]} \nonumber\\
& 	+
	(\text{bound state contribution}). \label{eq:6-3-2}
\end{align}
Observe that by tuning the parameters $\alpha_{\pm}$ to be $0$ or $\pi$, the kernel becomes the simpler form because, as mentioned in Section \ref{sec:RT}, in these spacial cases the phase shifts $\delta_{\pm}$ become independent of momentum: $\delta_{\pm} = 0$ for $\alpha_{\pm} = 0$ and $\delta_{\pm} = \pi$ for $\alpha_{\pm} = \pi$.
Thus there are $2\times 2= 4$ simpler cases, $(\alpha_{+}, \alpha_{-}) = (0, 0)$, $(\pi, \pi)$, $(0, \pi)$ and $(\pi, 0)$, which correspond to the Neumann-Neumann, Dirichlet-Dirichlet, Neumann-Dirichlet and Dirichlet-Neumann boundary conditions at $x=0$ and $L$.
These cases belong to the scale-independent subfamily of the $U(2)$ family of point interactions and the corresponding kernels are vastly studied in the literature \cite{Fulop:1999,Fulop:2003,Asorey:2007}, whose results coincide with \eqref{eq:6-3-2} as the above special cases.

It should be also pointed out here that the partial amplitude for $n=0$ in \eqref{eq:6-3-2} is nothing but the one-particle Feynman kernel evaluated on a half-line in the presence of a reflecting wall characterized by the boundary condition $\psi(0) + L_{+}\psi^{\prime}(0) = 0$ \cite{Farhi:1989,Fulop:2001b}.

\subsection{$\mathscr{P}$-symmetric subfamily} \label{subsec:parity}
Next consider the parity invariant point interactions, which include the famous $\delta$- and $\epsilon$-interactions.
The parity transformation on a circle is defined as $\mathscr{P}: x\mapsto L-x$.
It is known that the parity invariant point interactions are characterized by the matrix $U$ satisfying $U = \sigma_{1}U\sigma_{1}$ \cite{Fulop:1999,Fulop:2003}.
Such a matrix is obtained by choosing $e_{y} = e_{z} =0$ so that the S-matrix becomes $S^{(1)} = \mathrm{e}^{i\delta_{+}}\mathcal{P}_{+} + \mathrm{e}^{i(\delta_{-} + \pi)}\mathcal{P}_{-}$ for $e_{x} = 1$ and $S^{(1)} = \mathrm{e}^{i\delta_{-}}\mathcal{P}_{+} + \mathrm{e}^{i(\delta_{+} + \pi)}\mathcal{P}_{-}$ for $e_{x} = -1$, where $\mathcal{P}_{\pm}$ are the projection operators given by $\mathcal{P}_{\pm} = (\1\pm\sigma_{1})/2$.
In what follows we will consider the case $e_{x} = 1$.
The result for $e_{x} = -1$ can be obtained by changing $\delta_{\pm}$ to $\delta_{\mp}$.
By using the equation \eqref{eq:3-7}, the $n$-times scattering matrix is easily computed with the result
\begin{align}
S^{(n)}(k)
&= 	\frac{1}{2}
	\begin{pmatrix}
	\mathrm{e}^{in\delta_{+}(k)} + \mathrm{e}^{in(\delta_{-}(k) + \pi)} 	& \mathrm{e}^{in\delta_{+}(k)} - \mathrm{e}^{in(\delta_{-}(k) + \pi)} \\
	\mathrm{e}^{in\delta_{+}(k)} - \mathrm{e}^{in(\delta_{-}(k) + \pi)} 	& \mathrm{e}^{in\delta_{+}(k)} + \mathrm{e}^{in(\delta_{-}(k) + \pi)}
	\end{pmatrix}. \label{eq:6-4-1}
\end{align}
The Feynman kernel \eqref{eq:1-3} is therefore
\begin{align}
& 	K(x,T; x_{0},0) \nonumber\\
&= 	\sum_{n\in\mathbb{Z}}\int_{-\infty}^{\infty}\frac{\mathrm{d}p}{2\pi}
	\frac{\mathrm{e}^{in\delta_{+}(p)}}{2}
	\left\{
	\mathrm{e}^{iT\left[p\left(\frac{nL + x - x_{0}}{T}\right) - p^{2}\right]}
	+
	\mathrm{e}^{iT\left[p\left(\frac{(n+1)L - x - x_{0}}{T}\right) - p^{2}\right]}
	\right\} \nonumber\\
& 	+
	\sum_{n\in\mathbb{Z}}\int_{-\infty}^{\infty}\frac{\mathrm{d}p}{2\pi}
	\frac{\mathrm{e}^{in(\delta_{-}(p)+\pi)}}{2}
	\left\{
	\mathrm{e}^{iT\left[p\left(\frac{nL + x - x_{0}}{T}\right) - p^{2}\right]}
	-
	\mathrm{e}^{iT\left[p\left(\frac{(n+1)L - x - x_{0}}{T}\right) - p^{2}\right]}
	\right\} \nonumber\\
& 	+
	(\text{bound state contribution}). \label{eq:6-4-2}
\end{align}
Notice that the first and second lines are parity-even and -odd contributions respectively:
the former is invariant under the parity transformation $\mathscr{P}: x\mapsto L-x$, while the latter changes the sign.

\section{Conclusions and discussions} \label{sec:conclusion}
In this paper we studied the particle propagation on a circle in the presence of a single point interaction compatible with the conservation of the probability current, or the self-adjoint extension of the Laplace operator $-\mathrm{d}^{2}/\mathrm{d}x^{2}$ on $S^{1}\setminus\{0\}$.
We uncovered the classical trajectories for a quantum particle on $S^{1}\setminus\{0\}$, which consist of $2^{n}$ distinct paths for a particle scattered $n$-times from the point interaction (point scatterer).
We also illuminated deep connection between the scattering theory on $\mathbb{R}\setminus\{0\}$ and the spectral property of $S^{1}\setminus\{0\}$, which, in roughly speaking,  is summarized as the following correspondences:
\begin{center}
\begin{tabular}{l c l}
eigenvalues of S-matrix on $\mathbb{R}\setminus\{0\}$ 	& $\Leftrightarrow$ 	& energy spectrum of $S^{1}\setminus\{0\}$; \\
eigenvectors of S-matrix on $\mathbb{R}\setminus\{0\}$ 	& $\Leftrightarrow$ 	& energy eigenfunctions on $S^{1}\setminus\{0\}$.
\end{tabular}
\end{center}
We emphasize that the eigenvalues of the S-matrix depend only on the three parameters $\alpha_{+}$, $\alpha_{-}$ and $e_{x}$, whereas the eigenvectors depend on the full parameters of $U(2)$.
The reason will be explained as follows: Since the eigenvalues of the S-matrix on $\mathbb{R}\setminus\{0\}$ correspond to the energy spectrum of $S^{1}\setminus\{0\}$, we here explain why the energy spectrum of $S^{1}\setminus\{0\}$ does not depend on $e_{y}$ and $e_{z}$.
We first point out that the parity operator
\begin{align}
(\mathscr{\hat P}\psi)(x)
= 	\psi(L-x) \label{eq:7-1}
\end{align}
is well-defined on $S^{1}\setminus\{0\}$.
Let us then consider the following (singular) unitary transformation:
\begin{subequations}
\begin{align}
\psi(x)
&\stackrel{\mathcal{\hat U}}{\mapsto}
	{\tilde \psi}(x) = (\mathcal{\hat U}\psi)(x), \label{eq:7-2a}\\
H
&\stackrel{\mathcal{\hat U}}{\mapsto}
	{\tilde H} = \mathcal{\hat U}H\mathcal{\hat U}^{-1} = H, \label{eq:7-2b}
\end{align}
\end{subequations}
where
\begin{align}
\mathcal{\hat U}
:= 	\mathrm{e}^{i\beta\mathscr{\hat P}}, \quad
\beta\in\mathbb{R}, \label{eq:7-3}
\end{align}
which acts on the two-component vectors \eqref{eq:1-5} as follows:
\begin{subequations}
\begin{align}
{\vec \Psi}(x)
&\stackrel{\mathcal{\hat U}}{\mapsto}
	{\vec {\tilde \Psi}}(x)
	= 	\mathrm{e}^{i\beta\sigma_{1}}{\vec \Psi}(x), \label{eq:7-4a}\\
{\vec \Psi}^{\prime}(x)
&\stackrel{\mathcal{\hat U}}{\mapsto}
	{\vec {\tilde \Psi}^{\prime}}(x)
	=	\mathrm{e}^{i\beta\sigma_{1}}{\vec \Psi}^{\prime}(x). \label{eq:7-4b}
\end{align}
\end{subequations}
Since the unitary transformation leaves the Hamiltonian invariant, the energy spectrum remains the same but the boundary condition \eqref{eq:1-4} is changed as
\begin{align}
({\tilde U} - \1){\vec {\tilde \Psi}}(0_{+})
+ iL_{0}({\tilde U} + \1){\vec {\tilde \Psi}^{\prime}}(0_{+})
= 	{\vec 0}, \label{eq:7-5}
\end{align}
where
\begin{align}
{\tilde U}
= 	\mathrm{e}^{i\beta\sigma_{1}}U\mathrm{e}^{-i\beta\sigma_{1}}. \label{eq:7-6}
\end{align}
Thus, we found that the unitary transformation has an effect on $(e_{y}, e_{z})$ by a rotation of the angle $2\beta$, i.e.
\begin{align}
(e_{y}, e_{z})
&\stackrel{\mathcal{\hat U}}{\mapsto}
	(\cos(2\beta)e_{y} + \sin(2\beta)e_{z},
	-\sin(2\beta)e_{y} + \cos(2\beta)e_{z}). \label{eq:7-7}
\end{align}
This implies that the energy spectrum should depend on not $(e_{y}, e_{z})$ but their invariant $e_{y}^{2} + e_{z}^{2}$, which is identical to $1-e_{x}^{2}$.
Thus the spectrum depends only on the three parameters $(\alpha_{+}, \alpha_{-}, e_{x})$.

The main success of this work is the systematic description for a one-particle Feynman kernel on a circle with a point interaction.
The point is that we do not need any knowledge of the spectrum nor the complete set of energy eigenfunctions of the system (except for the bound states).
What we have to know is the classical trajectories of a particle and the one-particle scattering matrix.

We are left with a number of questions, however.
Let us close with a few comments on these issues:
\begin{enumerate}

\item \textit{More rigorous foundation of partial amplitude unitarity}.\\
We showed that the particle propagation scattered $n$-times from the point interaction should be weighted by the elements of $S^{(n)}(k)$.
As a direct consequence of the unitarity of the S-matrix, these weight factors satisfies the relation $\bigl|S^{(n)}_{\pm\pm}(k)\bigr|^{2} + \bigl|S^{(n)}_{\mp\pm}(k)\bigr|^{2} = 1$, which we proposed to call \textit{partial amplitude unitarity}.
As briefly discussed in Section \ref{sec:examples}, in the case of reflectionless subfamily of point interactions our partial amplitude unitarity and the Laidlaw-DeWitt theorem seem to be the same thing.
However, the theorem is essentially based on the homotopy theory so that it could not be applied in general to the other subfamily of point interactions.
We thought that the partial amplitude unitarity would provide a wider notion than the Laidraw-DeWitt theorem and should be derived from some fundamental properties of the Feynman kernel such as the unitarity $K(x,T; x_{0},0) = \int_{0}^{L}\!\mathrm{d}y K(x,T; y,t)K(y,t; x_{0},0)$.

Another related issue is an algebraic structure for the construction of classical trajectories for a particle on $S^{1}\setminus\{0\}$.
As mentioned before, the classical trajectories for a particle interacting $n$-times to the point interaction consist of $2^{n}$ distinct paths.
These trajectories are constructed from the more fundamental trajectories depicted in Figure \ref{fig:S-matrix} by gluing them under the multiplication rule of the matrix $S^{(1)}$.
This fact implies that, in spite of the presence of a point singularity, it might be possible to introduce some kind of notion analogous to the fundamental group to the space $S^{1}\setminus\{0\}$.
However, we have no idea to treat these problems.

\item \textit{Path-integral representation for the kernels}.\\
The Feynman kernels derived in this paper have the forms which will be obtained after performing the path-integration.
It is very interesting to investigate its path-integral representation.
As mentioned in the Introduction, boundary conditions are treated unambiguously in operator formalism by von Neumann's theory of self-adjoint extension.
However, in path-integral formalism we do not know \textit{a priori} what kind of trajectories we should integrate over and what kind of ``classical action'' we should adopt as a weight.
Indeed, even in the system of a free particle in an infinitely deep well potential, the ``classical action'' we should adopt as a path-integral weight includes the so-called \textit{topological term} that localizes at the boundaries \cite{Chaichian:2001}.
Furthermore, this topological term is proportional to the Planck constant so that it is no longer classical.

To that aim we have to consider the limit $N\to\infty$ of the equation:
\begin{align}
\langle x|\mathrm{e}^{-iHT}|x_{0}\rangle
= 	\left[\prod_{n=1}^{N-1}\int_{0}^{L}\!\!\!\mathrm{d}x_{n}\right]
	\langle x|\mathrm{e}^{-iH\delta t}|x_{N-1}\rangle
	\cdots
	\langle x_{1}|\mathrm{e}^{-iH\delta t}|x_{0}\rangle,
\end{align}
where $\delta t:= T/N$.
We already know the exact form of each partial amplitude $\langle x_{i+1}|\mathrm{e}^{-iH\delta t}|x_{i}\rangle$.
In order to evaluate the right hand side we have to tackle with the $N-1$ products of both translational invariant and variant classes of infinite sums.
We would like to report this issue elsewhere.
\end{enumerate}

\section*{Acknowledgment} \label{sec:Acknowledgment}
S.O. would like to thank Y. Adachi and K. Sakamoto for valuable conversations and the latter also for the collaborations on several issues relevant in this paper.
M.S. would like to thank K. Takenaga and T. Nagasawa for useful discussions. This work is supported in part by the Grant-in-Aid for Scientific Research (No.18540275 (M.S.)) by the Japanese Ministry of Education, Science, Sports and Culture.


\end{document}